\documentclass[a4paper]{jpconf}
\usepackage{graphicx}
\usepackage{citesort}
\usepackage{color}

\begin{document}
\title{Observation of an Efimov resonance in an ultracold mixture of atoms and weakly bound dimers}

\author{S. Knoop$^{1,}$\footnote[3]{Present address: Kirchhoff-Institut f\"ur Physik, Universit\"at Heidelberg, Germany}, F. Ferlaino$^1$, M. Berninger$^1$, M. Mark$^{1,}$\footnote[4]{Present address: Swinburne University of Technology, Melbourne, Australia}, H.-C. N\"{a}gerl$^1$ and R. Grimm${^{1,2}}$}

\address{$^1$ Institut f\"ur Experimentalphysik and Zentrum f\"ur Quantenphysik, Universit\"at Innsbruck, 6020 Innsbruck, Austria}

\address{$^2$ Institut f\"ur Quantenoptik und Quanteninformation, \"Osterreichische Akademie der Wissenschaften, 6020 Innsbruck, Austria}

\ead{knoop@kip.uni-heidelberg.de}

\begin{abstract}
We discuss our recent observation of an atom-dimer Efimov resonance in an ultracold mixture of Cs atoms and Cs$_2$ Feshbach molecules [Nature Phys. 5, 227 (2009)]. We review our experimental procedure and present additional data involving a non-universal $g$-wave dimer state, to contrast our previous results on the universal $s$-wave dimer. We resolve a seeming discrepancy when quantitatively comparing our experimental findings with theoretical results from effective field theory.
\end{abstract}

\section{Introduction}

Atomic collisions play a central role in the field of ultracold quantum gases \cite{Burnett2002qeo}. Elastic collisions are required for evaporative and sympathetic cooling in order to enter the ultracold regime and to reach Bose-Einstein condensation or degeneracy in Fermi gases, whereas inelastic collisions often give rise to trap loss and therefore limit the lifetime of these ultracold gases. Only $s$-wave collisions are usually relevant in the ultracold regime, because all partial-wave contributions are suppressed. Therefore the interaction between two atoms is described by a single parameter, namely the $s$-wave two-body scattering length $a$. The presence of Feshbach resonances makes it possible to tune $a$ by means of a magnetic field, resulting in an unprecedented control of the interaction properties \cite{Chin2008fri}.

The tunability of the \emph{two-body} interaction allows to study \emph{few-body} quantum phenomena with ultracold gases in a rather new and remarkable way. In particular, the universal regime characterized by large $|a|$ can be reached. Here details of the underlying short-range interaction are unimportant and few-body properties become universal as they only depend on $a$ \cite{Braaten2006uif}. For atomic systems universality requires $a$ to be much larger than the van der Waals length $r_{\rm vdW}$ \cite{Kohler2006poc,Chin2008fri}.
A paradigm of universality in three-body quantum physics is the Efimov effect. In the 1970's Vitaly Efimov predicted that for $|a|$$\rightarrow$$\infty$ an infinite number of trimer states exists, even in the absence of weakly bound two-body states \cite{Efimov1970ela,Efimov1971wbs}. An important consequence is that low-energy scattering properties of three-body systems are affected at those values for the scattering length at which Efimov states are resonantly coupled to the relevant dissociation thresholds \cite{Efimov1979lep}.

The first experimental evidence of an Efimov state was found in an ultracold gas of Cs atoms \cite{Kraemer2006efe}, observed as a giant resonance in the three-body recombination loss rate at large negative $a$ \cite{Esry1999rot,Braaten2001tbr}. Since then, signatures of Efimov states \cite{Nielsen1999ler,Esry1999rot,Bedaque2000tbr,Braaten2001tbr} have been found in atomic gases of $^{39}$K \cite{Zaccanti2009ooa} and $^{7}$Li \cite{Gross2009oou}, which like the Cs experiment represent systems of identical bosons, as in the original problem considered by Efimov. Systems of non-identical particles can also be relevant for the Efimov effect \cite{Amado1972eea,Efimov1973elo}, and signatures have been found in a three-component $^6$Li gas \cite{Ottenstein2008cso,Huckans2009tbr,Wenz2009aut} and an ultracold mixture of $^{41}$K and $^{87}$Rb \cite{Barontini2009ooh}.

All these experiments rely on measuring three-body recombination loss in an ultracold, trapped sample of atoms.
However, an essential part of the Efimov scenario involves the interaction of the trimers near the atom-dimer threshold, which can be directly probed via relaxation loss in a mixture of atoms and weakly bound dimers \cite{Nielsen2002eri,Braaten2004edr}. Recently, we have prepared mixtures of Cs atoms and weakly bound Cs$_2$ Feshbach molecules and found an atom-dimer Efimov resonance \cite{Knoop2009ooa}. Here we will review our experimental procedure and present new experimental data showing that for a non-universal dimer the resonance feature is absent. We will compare our data with recent effective field theory calculations \cite{Braaten2009rdr,Helfrich2009rad}, which resolves the discrepancy with Ref.~\cite{Braaten2007rdr}, and allow for a reinterpretation of our data regarding the quantitative comparison with universal theory.

\section{Efimov scenario and its impact on ultracold gases}

\begin{figure}
\begin{center}
\includegraphics[width=4.5in]{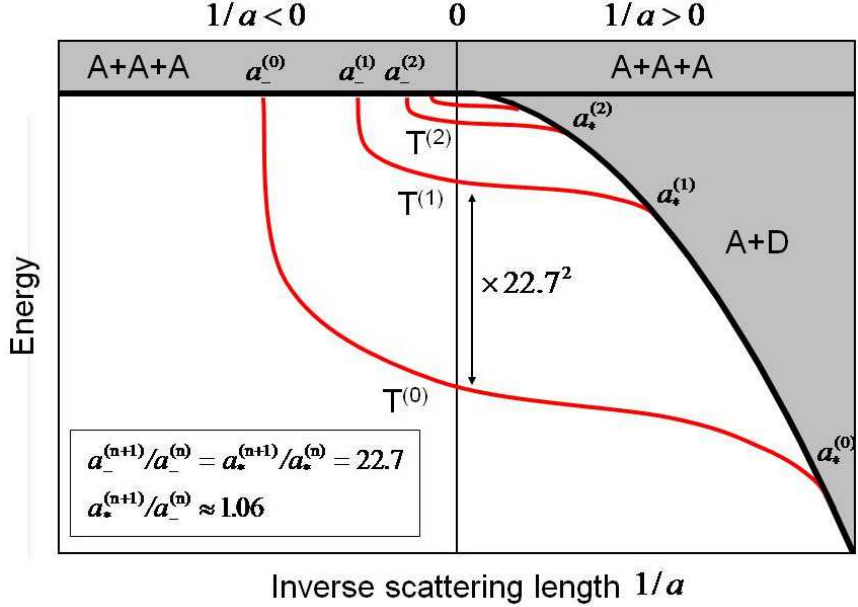}
\caption{Efimov's scenario, illustrating the appearance of an infinite series of weakly bound Efimov trimer states T$^{(n)}$ for resonant two-body interaction, exhibiting a logarithmic-periodic pattern with universal scaling factors $22.7$ and $22.7^2$ for the scattering length and the binding energy, respectively. The binding energy is plotted as a function of the inverse two-body scattering length $1/a$. The gray regions indicate the scattering continuum for three atoms (A+A+A) and for an atom and an universal dimer (A+D). The scattering lengths at which the trimers are degenerate with the three-atom and atom-dimer threshold are labeled as $a^{(n)}_-$ and $a^{(n)}_*$, respectively. To illustrate the series of Efimov states, the universal scaling factor is artificially set to 2.}\label{efimovscenario}
\end{center}
\end{figure}

In figure \ref{efimovscenario} the Efimov scenario is depicted, showing Efimov's predicted geometrical spectrum of trimer states in the limit of large $a$. For $a$$<$0 Efimov states cross the atomic threshold at $a^{(n)}_-$. For $a$$>$0 a universal weakly bound $s$-wave dimer state exists with binding energy $E_{\rm b} = \hbar/(m a^2)$, $m$ being the atomic mass. The trimers merge into the corresponding atom-dimer threshold at $a^{(n)}_*$. For three identical bosons in the universal limit the positions $a^{(n)}_-$ and $a^{(n)}_*$ show both a scaling by a factor $22.7$ for subsequent Efimov states, whereas the binding energy scales with $22.7^2=515$. Furthermore, in the ideal Efimov scenario a universal connection between $a$$<$0 and $a$$>$0 exists via $a^{(n+1)}_*\approx1.06\times |a^{(n)}_-|$ \cite{Braaten2006uif,Gogolin2008aso}.

The Efimov scenario impacts three-body recombination in ultracold atomic samples in two distinct ways: enhancement of the recombination loss rate coefficient at $a^{(n)}_-$, where the resonant coupling of three atoms to an Efimov state opens up additional decay channels \cite{Esry1999rot}, and minima at $a^{(n)}_+=\sqrt{22.7}|a^{(n)}_-|\approx a^{(n)}_*/(1.06\sqrt{22.7})$, which are explained by destructive interference between two pathways that both lead to recombination into the weakly bound dimer \cite{Nielsen1999ler,Esry1999rot}. So far, several of these Efimov features have been found \cite{Kraemer2006efe,Zaccanti2009ooa,Gross2009oou,Ottenstein2008cso,Huckans2009tbr,Barontini2009ooh}, including two successive loss minima \cite{Zaccanti2009ooa} and a universal connection between a resonance and a minimum  \cite{Gross2009oou}.

The appearance of an Efimov trimer at the atom-dimer threshold at $a^{(n)}_*$ is predicted to manifest itself in a resonant enhancement of atom-dimer relaxation \cite{Nielsen2002eri,Braaten2004edr}. In the context of ultracold gases, the universal weakly bound dimers, which have binding energies much smaller than $E_{\rm vdW}=\hbar/(m r_{\rm vdW}^2)$, are connected to vibrationally highly excited molecular states. Collisions with atoms may lead to relaxation to more deeply bound dimer states. In such an inelastic collision both the atom and the dimer are lost from the trap, as the corresponding release of energy generally exceeds the trap depth. The particle loss is described by the rate equation
\begin{equation}\label{moleculeequation}
\dot{n}_{\rm D}=\dot{n}_{\rm A}=-\beta n_{\rm D} n_{\rm A},
\end{equation}
where $n_{\rm D}$($n_{\rm A}$) is the molecular (atomic) density and $\beta$ denotes the loss rate coefficient for atom-dimer relaxation. Especially if the ultracold sample contains much more atoms than dimers, the loss of dimers is dominated by atom-dimer relaxation and loss due to dimer-dimer relaxation plays a minor role. Typically the lifetimes of a trapped dimer sample in ultracold atom-dimer mixtures are short ($\ll$ 1~s). However, as long as these lifetimes are much larger than the time required to prepare the atom-dimer mixture, the inelastic loss rate $\beta$ can be measured.

Within the framework of effective field theory Braaten and Hammer derived a universal formula for relaxation into a deeply-bound dimer state for a system of identical bosons at zero temperature \cite{Braaten2004edr}:
\begin{equation}\label{efimovformula}
\beta= \frac{20.3\sinh(2\eta_*)}{\sin^2\left[s_0\ln(a/a^{(n)}_*)\right]+\sinh^2\eta_*} \frac{\hbar a}{m},
\end{equation}
where $\eta_*$ corresponds to the decay parameter, related to the lifetime of the trimer state. Thus, the Efimov scenario is expected to show up as loss resonances in mixtures of atoms and weakly bound dimers, i.e.\ atom-dimer Efimov resonances, on top of a linear $a$ scaling of $\beta$ (see also Ref.~\cite{DIncao2005sls}).

Refs.~\cite{Zaccanti2009ooa,Barontini2009ooh} have reported on narrow loss features at $a$$>$0 in pure atomic samples, which they have interpreted as atom-dimer Efimov resonances, assuming a multiple collision process. The interpretation is as follows: the weakly bound dimer formed in the initial three-body recombination process kicks out several atoms by elastic collisions before the dimer undergoes an inelastic collision and leaves the trap. This can only occur at $a^{(n)}_*$ where the atom-dimer scattering length is very large. Still, very high densities are required for this process to be observable, and furthermore it is very difficult to extract $\beta$ from these indirect observations, hindering a quantitative verification of the universal prediction.

\section{Experiment}

\begin{figure}
\begin{center}
\includegraphics[width=4in]{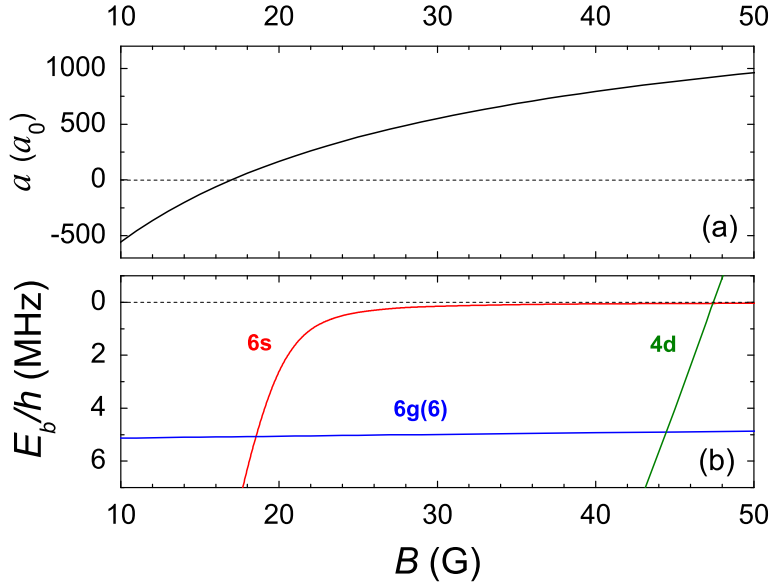}
\caption{The magnetic field dependence of the scattering length $a$ (a) and near-threshold dimer spectrum (b) of Cs, considering two atoms in their lowest hyperfine sublevel. The narrow $d$- and $g$-wave resonances \cite{Mark2007sou}, including the one at 48 G, are not shown in (a). We have depicted only the three molecular states that are relevant to this work: the $s$-wave dimer state $6s$ that becomes the universal weakly bound state for $B>20$ G, and two non-universal $d$- and $g$-wave states, labeled as $4d$ and $6g(6)$. For the labeling of the molecular states, see Ref.~\cite{Mark2007sou}. The intersections between these dimer states are in reality avoided crossings, which can be used to populate these dimer states \cite{Mark2007sou,Ferlaino2009ufm}.}\label{scatteringenergy}
\end{center}
\end{figure}

We have performed the experiment starting with Cs atoms in the lowest hyperfine sublevel, with hyperfine and
projection quantum numbers $F=3$ and $m_F=3$. Here $a$ shows a pronounced dependence on the magnetic field in the low-field region below 50\,G; see figure~\ref{scatteringenergy}(a). Over a wide range, $|a|$ is very large and exceeds $r_{\rm vdW} \simeq 100\,a_0$, where $a_0$ is Bohr's radius. A near-universal weakly bound $s$-wave dimer state with $E_{\rm b}<E_{\rm vdW} \simeq h\times2.7$~MHz exists for $B>20$ G. In figure~\ref{scatteringenergy}(b) this dimer state is labeled as $6s$. For $B<20$ G this state bends towards larger $E_{\rm b}$ and becomes non-universal \cite{Mark2007sou}. In this paper we will also consider atom-dimer relaxation loss involving the $g$-wave $6g(6)$ state, which is non-universal for all magnetic fields. Note that we have recently studied inelastic dimer-dimer collisions involving universal and non-universal dimer states, including the $6g(6)$ state \cite{Ferlaino2009cou}.

We prepare an ultracold atomic sample in a crossed-beam optical dipole trap, which consists of two 1064-nm laser beams with
waists of about 250~$\mu$m and 36~$\mu$m, the latter being spatially modulated by an AOM to obtain a tunable elliptical beam \cite{Ferlaino2008cbt}. Part of the atomic ensemble is converted into dimers by means of Feshbach association \cite{Kohler2006poc,Ferlaino2009ufm} using a 200-mG wide Feshbach resonance at 48~G \cite{Mark2007sou,Ferlaino2008cbt}, which is induced by the $4d$ state (see figure~\ref{scatteringenergy}). Except for the application of optical lattices, Feshbach association can never be adiabatic for bosons because of strong loss at the Feshbach resonance. Therefore one always ends up with an imbalanced mixture of the atoms and dimers. For our lowest temperatures of 30~nK we obtain a mixture of about $3\times10^{4}$ atoms and $4\times10^{3}$ dimers. Even though the association proceeds via a $d$-wave Feshbach resonance, the weakly bound $s$-wave state is easily accessible due to an avoided crossing between the $6s$ and the $4d$ states. In general, avoided crossings between the molecular states allow for the population of states that are not directly coupled to the atomic threshold \cite{Mark2007sou,Ferlaino2009ufm}.

After preparation of the mixture we ramp to a certain magnetic field and wait for a variable storage time. Then we switch off the trap and let the sample expand before ramping back over the 48-G resonance to dissociate the molecules, after which standard absorption imaging is performed. During the expansion a magnetic field gradient is applied to spatially separate the atomic and molecular cloud (Stern-Gerlach separation) \cite{Herbig2003poa}. In this way we simultaneously monitor the number of remaining atoms and dimers.

\begin{figure}
\begin{center}
\includegraphics[width=5in]{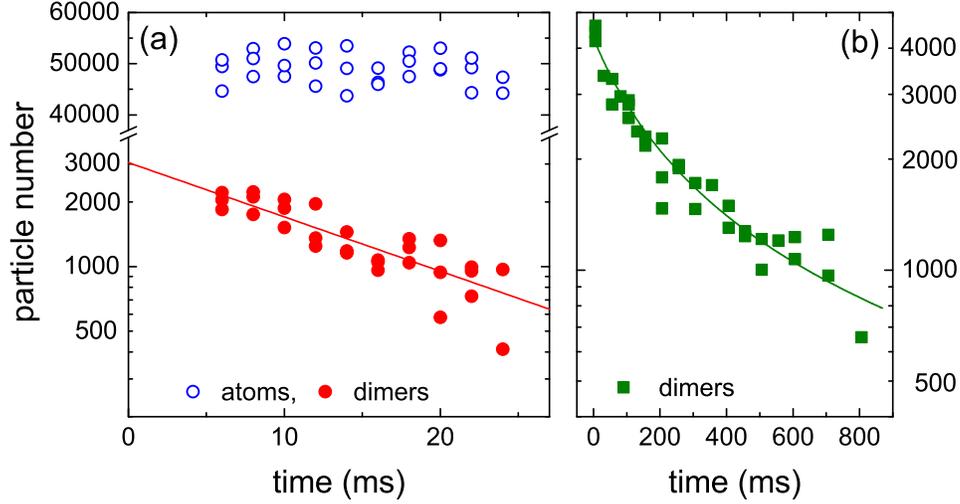}\label{lifetimes}
\caption{Time evolution of the number dimers at 35~G, comparing the dimer lifetime in an atom-dimer mixture consisting of Cs atoms in the lowest hyperfine sublevel and Cs Feshbach molecules in the $6s$ state (a) and a pure $6s$ dimer sample (b). In (a) the loss of dimers can be fitted with an exponential decay curve with a $1/e$ lifetime proportional to $\beta^{-1}$, as the atom number greatly exceeds the dimer number and loss due to the dimer-dimer relaxation can be neglected, as can be immediately seen from the much longer timescale over which the pure dimer sample decays.}
\end{center}
\end{figure}

We measure the atom-dimer relaxation loss rate $\beta$ by recording the time evolution of the dimer number $N_{\rm D}$ and atom number $N_{\rm A}$. A typical loss measurement is shown in figure~\ref{lifetimes}(a). When $N_{\rm A}$ greatly exceeds $N_{\rm D}$ and dimer-dimer relaxation loss is negligible, $N_{\rm D}$ shows an exponential decay with a $1/e$ lifetime that is proportional to $\beta^{-1}$. For comparison, a lifetime measurement of a pure dimer sample is presented in figure~\ref{lifetimes}(b), showing that the loss rate due to dimer-dimer relaxation is indeed much smaller. A detailed description of our data analysis can be found in Ref.~\cite{Knoop2009ooa}.

\section{Results}

\begin{figure}
\begin{center}
\includegraphics[width=6in]{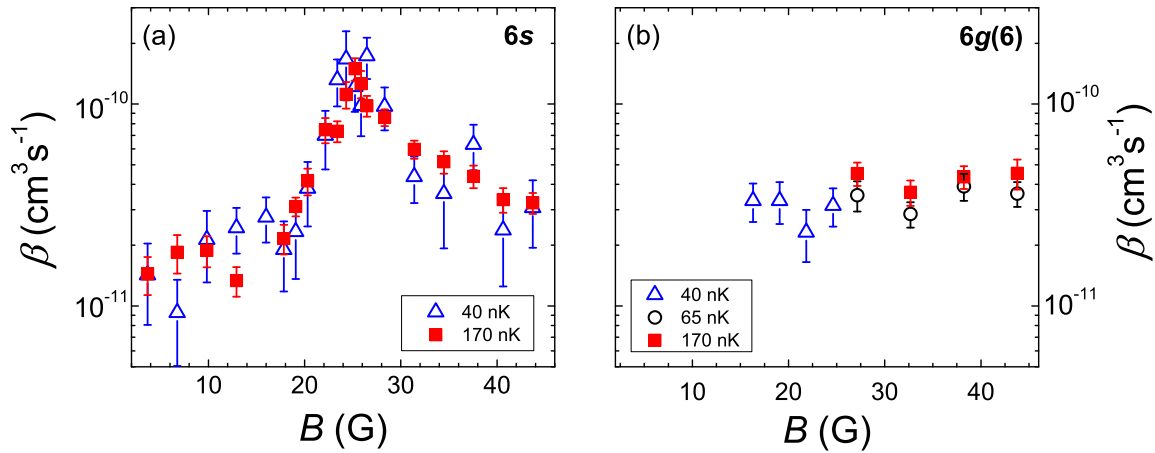}\label{AD_magneticfield}
\caption{The loss rate coefficient $\beta$ for atom-dimer relaxation involving the $6s$ and $6g(6)$ dimer states are shown in panel (a) and (b), respectively, as a function of the magnetic field $B$. The error bars on $\beta$ contain all statistical uncertainties (one standard deviation) from the fit of the time evolution as well as the trap frequencies and the temperature measurements.}
\end{center}
\end{figure}

We have measured $\beta$ for the $6s$ and $6g(6)$ dimer states for magnetic fields up to about 45~G, and the results are shown in figure~\ref{AD_magneticfield}(a) and (b), respectively. For the $6s$ state we observe a strong loss resonance where the loss rate shows an increase of one order of magnitude. The resonance is located at about 25\,G where the $6s$ state represents the weakly bound dimer state in the near-universal regime. In contrast, the $6g(6)$ state, which has no universal character, shows no magnetic field dependence. This shows that the resonance is a unique feature of the weakly bound $s$-wave dimer state.

\begin{figure}
\begin{center}
\includegraphics[width=4in]{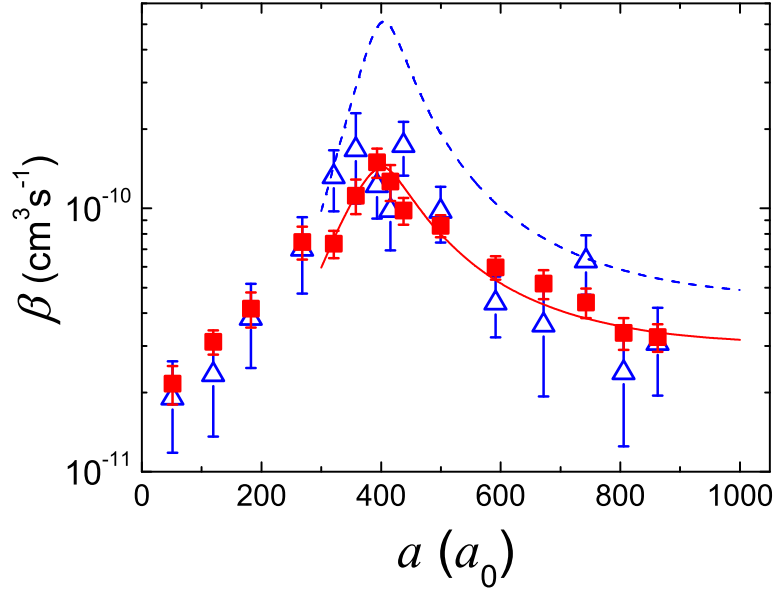}
\caption{The loss rate coefficient $\beta$ for atom-dimer relaxation involving the $6s$ dimer state, showing the same data as in figure~\ref{AD_magneticfield}(a), but here as function of $a$ and only for $a$$>$0. The measurements are taken at temperatures of 40(10)~nK (blue open triangles) and 170(20)~nK (red closed squares). The curves are taken from Ref.~\cite{Helfrich2009rad}, calculated within the effective field theory framework and using a Bose-Einstein thermal averaging procedure (blue dashed curve 40 nK, red solid curve 170 nK). The parameters $a_*$ and $\eta_*$ are obtained from a fit to the 170 nK data, resulting in $a_*=397\,a_0$ and $\eta_*=0.034$.}\label{AD_scatteringlength}
\end{center}
\end{figure}

In figure~\ref{AD_scatteringlength} the $6s$ data is presented as function of $a$, using the well-known $B$-dependence of $a$ as shown in figure~\ref{scatteringenergy}(a), and only the $a$$>$0 region is displayed. The absence of a temperature dependence suggested the direct application of the zero-temperature result, equation~(\ref{efimovformula}), to fit our data in the region where $a>r_{\rm vdW}$. However, for equation~(\ref{efimovformula}) to fit our data requires the prefactor of 20.3 to be a free fit parameter. From our 170 nK data we obtain $a_*=367(13)\,a_0$, $\eta_*=0.30(4)$, and the prefactor to be 2.0(2) \cite{Knoop2009ooa}, i.e.\ one order of magnitude smaller than theory predicted.

The modification the universal prediction was motivated by the results of Ref.~\cite{Braaten2007rdr}. These calculations showed a strong temperature dependence and much higher loss rate coefficients than observed in our experiment, which was interpreted as a manifestation of the unitarity limit in combination with a too large zero temperature loss rate coefficient. However, Ref.~\cite{Braaten2007rdr} suffered from a computational error in the thermal averaging and corrected calculations show much smaller loss rates at finite temperature \cite{Braaten2009rdr}. Furthermore, Hammer and Helfrich \cite{Helfrich2009rad} reported on more advanced calculations compared to Ref.~\cite{Braaten2009rdr}, by using the full effective field theory results for the atom-dimer phase shifts instead of the effective range expansion and performing thermal averaging using Bose-Einstein instead of Boltzmann distributions. A detailed comparison between the calculations of Refs.~\cite{Braaten2009rdr} and \cite{Helfrich2009rad} shows only small differences \cite{Helfrich2009private}.

The main results of the calculations of Ref.~\cite{Helfrich2009rad} are included in figure~\ref{AD_scatteringlength}. A fit to the 170 nK data, for $a>300 a_0$ and with only $a_*$ and $\eta_*$ as free parameters, describes the data very well and results in $a_*=397\,a_0$ and $\eta_*=0.034$. Besides the slightly larger $a_*$ than obtained in Ref.~\cite{Knoop2009ooa}, $\eta_*$ is found to be much closer to 0.06, which was the value found for the three-body recombination resonance at $a_-=-850a_0$ \cite{Kraemer2006efe}. With the same $a_*$ and $\eta_*$ values obtained from the fit of the 170 nK data, the calculations overestimate the 40 nK data by a factor 2.

The results of Ref.~\cite{Helfrich2009rad} allow for a reinterpretation of our data. The 170 nK data agrees very well with the universal description of an atom-dimer Efimov resonance. However, according to theory the 40 nK measurements should have shown higher loss rate coefficients compared to the 170 nK data, but we do not observe such a temperature dependence. The origin of this discrepancy is presently not clear. However, we note that the optical trap required to obtain a temperature of 40 nK was much weaker than the one for 170 nK, and therefore more susceptible to imperfections, which could lead to a systematic error in determining the properties of the trap and therefore the extracted loss rates.

Finally we consider the universal connection between $a$$>$$0$ and $a$$<$$0$, given by $a^{(n+1)}_*/|a^{(n)}_-|\approx1.06$ in the ideal Efimov scenario \cite{Braaten2006uif,Gogolin2008aso}. With the observation of $a_-=-850\,a_0$ \cite{Kraemer2006efe} and the result of $a_*=397\,a_0$ \cite{Knoop2009ooa,Helfrich2009rad} we obtain $a_*/|a_-|=0.47$, which is significantly smaller than the value 1.06. This could be, at least partly, explained by finite-range corrections, which are particularly important for low-lying Efimov states \cite{Hammer2007erc,Thogersen2008upo,DIncao2009iit,Platter2009rct}. An alternative explanation would be a change in the three-body parameter between the $a$$<$$0$ and $a$$>$$0$ regions, which could occur in our case because these regions are connected via a zero crossing instead of a pole in $a$ (see figure~\ref{scatteringenergy}(a)) \cite{DIncao2009iit}. This is still an open issue that requires further studies, both experimentally and theoretically.

\section{Conclusions}

We have studied relaxation loss in an ultracold atom-dimer mixture consisting of Cs atoms and Cs$_2$ Feshbach molecules. In the region of large and positive $a$ we have observed a resonance in the relaxation loss rate for the weakly bound $s$-wave dimer. When we change the dimer state to that of a non-universal $g$-wave state this resonant feature is absent and a constant relaxation rate is observed in the same magnetic field region.

Recent work \cite{Braaten2009rdr,Helfrich2009rad} show the necessity of thermal averaging in the calculation of relaxation loss around the atom-dimer Efimov resonance, even at our ultralow temperatures. This means that the zero-temperature result cannot be directly applied to fit the data, as was done in Ref.~\cite{Knoop2009ooa}. This new insight reveals that our data at 170 nK do show quantitative agreement with the universal description of an atom-dimer Efimov resonance, whereas the loss rate coefficients of our 40 nK data are a factor of 2 too low. Our result demonstrates that atom-dimer relaxation measurements provide complementary information on Efimov physics to that obtained by three-body recombination in pure atomic gases.

\ack
We thank H.-W. Hammer and K. Helfrich for helpful discussions and for providing their calculations. We acknowledge support by the Austrian Science Fund (FWF) within SFB 15 (project part 16). S.~K.\ is supported within the Marie Curie Intra-European Program of the European Commission. F.~F.\ is supported within the Lise Meitner program of the FWF.

\section*{References}

\providecommand{\newblock}{}

\end{document}